\documentclass[12pt]{iopart}

\usepackage{graphicx}

\expandafter\let\csname equation*\endcsname\relax
\expandafter\let\csname endequation*\endcsname\relax
\usepackage{mathtools}
\usepackage{amssymb}
\usepackage{hyperref}
\usepackage{dsfont}
\usepackage{xcolor}

\usepackage{float} 
\usepackage[section]{placeins} 
\usepackage{comment}
\usepackage[normalem,normalbf]{ulem} 

\newcommand{\bra}[1]{\left < {#1} \right | }
\newcommand{\ket}[1]{\left | {#1} \right > }
\newcommand{\braket}[2]{\left < {#1} \middle | {#2} \right > }

\newcommand\numberthis{\addtocounter{equation}{1}\tag{\theequation}}

\begin{document}
\title[Optimization of Ab-Initio Based Tight-Binding Models]{Optimization of Ab-Initio Based Tight-Binding Models}
\author{Henrik Dick and Thomas Dahm}
\address{Bielefeld University, Physics Department, Postfach 100131, D-33501 Bielefeld, Germany}
\ead{\mailto{hdick@physik.uni-bielefeld.de}}
\vspace{10pt}
\begin{indented}
\item[]\today
\end{indented}

\begin{abstract}
\label{Abstract}

The electronic structure of solids can routinely be calculated by standard methods like density functional theory. However, in complicated situations like interfaces, grain boundaries or contact geometries one needs to resort to more simplified models of the electronic structure. Tight-binding models are using a reduced set of orbitals and aim to approximate the electronic structure by short range hopping processes. For example, maximally localized Wannier functions are often used for that purpose. However, their accuracy is limited by the need to disentangle the electronic bands. Here, we develop and investigate a different procedure to obtain tight-binding models inspired by machine-learning techniques. The model parameters are optimized in such a way as to reproduce ab-initio band structure data as accurately as possible using an as small as possible number of model parameters. The procedure is shown to result in models with smaller ranges and fewer orbitals than maximally localized Wannier functions but same or even better accuracy. We argue that such a procedure is more useful for automated construction of tight-binding models particularly for large-scale materials calculations.\\

\end{abstract}
\section{Introduction}\label{Introduction}

A standard method to calculate electronic properties of a crystalline solid is ab-initio band structure calculation based on density functional theory (DFT). In situations where periodicity is broken, like defects, grain boundaries, interfaces between materials, or complex geometries in electronic devices, ab-initio methods quickly become intractable, however. Ab-initio based tight-binding models have been suggested as a possible simplification that on one hand allows treating much larger and inhomogeneous systems but on the other hand retains the quantum mechanical nature of the bonding in a reduced local real space description.\cite{Goringe_1997,Seifert1998,Frauenheim2020} Tight-binding models usually start from a limited set of orbitals at the atomic sites which are used to represent the Hamiltonian of a material in a real space basis.\cite{SlaterKoster} As the matrix elements fall off with distance quickly, tight-binding models are usually restricted in range. The electronic structure can then be calculated by diagonalization of small matrices representing the Hamiltonian. Tight-binding models for specific materials are not unique as they depend on the chosen real space basis. Different kinds of basis sets can be used, like linear combination of atomic orbitals (LCAO),\cite{SlaterKoster,Papaconstantopoulos_2003,Papaconstantopoulos_book,TBPao} linear muffin-tin orbitals (LMTO), \cite{AndersenNMTO} or Wannier functions, \cite{MazariMLWF,SouzaMLWF,MazariReview} for example. The choice of the basis set influences the range of the tight-binding model that is necessary to accurately represent the electronic structure.

There are two main problems associated with the choice of localized sets of orbitals that have been pointed out in the past: one concerns the automation and the other the accuracy of the models. \cite{Gresch2018,Wang2021} Before a tight-binding model is constructed from a DFT bandstructure calculation a choice needs to be made, which localized orbitals at which positions are taken. Also, the target bands in the band structure that should be represented by the tight-binding model need to be chosen. These choices usually require human control, which is problematic for automated high-throughput construction of tight-binding models. The other problem is related to the projectability of the bands to the chosen set of localized orbitals. \cite{TBPao} In metallic systems, for example, it is necessary to disentangle the bands before a projection to the localized basis is made, \cite{MazariReview,Wannier90} which introduces a systematic error into the tight-binding model and limits its accuracy. This systematic error can be reduced by choosing a larger basis set. However, this requires to repeat the band structure calculation with a larger number of bands and a projection onto a larger set of localized orbitals, which quickly becomes very time-consuming.

A possible way out of these problems has been suggested by Wang et al.~\cite{Wang2021} The idea is not to specify the atomic orbitals from the beginning but instead let a machine-learning procedure find an optimum number of orbitals. This is achieved with a tight-binding Hamiltonian construction neural network (TBHCNN). The parameters of the network are trained using standard machine learning techniques in such a way as to represent the band structure as accurately as possible with the smallest possible number of parameters. The number of orbitals is gradually increased until a desired accuracy is obtained. On one hand this allows systematic control of accuracy on the other hand such a procedure does not require human control. The feasibility of this procedure was demonstrated for quasi one-dimensional systems (nanoribbons) as well as two-dimensional systems.

In this work we propose a method similar in spirit to the TBHCNN method,\\
however using conjugate gradient optimization of the tight-binding parameters based on first order perturbation theory. While for classical deep neural networks such a method usually does not significantly improve convergence due to the nonlinearities of the network, it turns out that it allows much faster convergence for bandstructures, as it can approach the optimized set of parameters in much larger steps. Also, a preconditioner can be used to further improve convergence.

In section \ref{Tight-binding models} we will briefly introduce tight-binding models along with an analysis of the computational cost needed to diagonalize them. In section \ref{Optimization technique} we introduce the optimization procedure we propose. As an illustration, in section \ref{Application to selected materials} we apply our procedure to selected three-dimensional materials. Section \ref{Wavefunctions and Orbitals} contains a discussion of the computation of the corresponding localized orbitals. A comparison with the standard code \emph{wannier90} is made in section \ref{Comparison with Wannier90}. As an application of our tight-binding models, the calculation of a Fermi surface is illustrated in section \ref{Fermi Surface}. Conclusions are found in section \ref{Conclusion}.

\section{Tight-binding models}\label{Tight-binding models}

In practice the bandstructure of a crystalline solid is described by a countable basis of Bloch-functions. They can be written as superposition of orbitals on a lattice:\\ \begin{align}  
\Psi_{n,\vec{k}}^{\sigma}(\vec{r})=\frac{1}{\sqrt{ N }}\sum_{\vec{R}}e^{i\vec{k}\vec{R}}\phi_{n}^{\sigma}(\vec{r}-\vec{R})  
\end{align} \\
where $\phi_{n}^{\sigma}$ represent a set of localized orbitals with spin $\sigma$. In the following, the spin will be absorbed into the quantum number $n$. The Hamiltonian can be represented using this basis \cite{SlaterKoster}\\
$$  
\left< \Psi_{n,\vec{k}}\middle| H \middle|\Psi_{m,\vec{k}} \right> = \sum_{\vec{R}}\exp(i\vec{k}(\vec{R}-\vec{R}'))\int \overline{ \phi_{n} }(\vec{r}-\vec{R}')\,H\,\phi_{m}(\vec{r}-\vec{R}) \, d^{3}r  
$$\\
Since the sum is taken over the whole lattice, $\vec{R}'$ can be set arbitrarily without changing the result. It is set to zero in the following. In this way the Hamiltonian has been transformed to the tight-binding form where $H_{nm}^{\vec{R}}$ are the tight-binding parameters.\\ \begin{align}  
\boxed{ \left< \Psi_{n,\vec{k}}\middle| H \middle|\Psi_{m,\vec{k}} \right>=\sum_{\vec{R}}e^{i\vec{k}\vec{R}}H^{\vec{R}}_{nm} }\label{eq:tb}  
\end{align} \\
The idea of tight-binding is, that the orbitals are localized and the integrals, which represent $H_{nm}^{\vec{R}}$, are therefore small for large distance $|\vec{R}|$. We consider finite tight-binding models, where the basis has been cut at some point. If the cut bands don't interact with the kept bands, the band energies will be preserved when cutting off the basis. There are different approaches to cutting the basis, depending on whether symmetry should be preserved or not. Slater-Koster tight-binding models \cite{SlaterKoster} use symmetrically complete sets of atomic orbitals and therefore have known transformation properties under rotation and translation of the lattice. Models based on maximally localized Wannier function can have any cut basis and are usually not symmetric. Symmetry adapted maximally localized Wannier functions have been studied in \cite{symwannier90}.

The bandstructure can be computed by diagonalizing the Hamiltonian. Note, that the basis doesn't need to be orthogonal, so a generalized eigenvalue problem has to be solved.\\
$$  
H_{nm}(\vec{k}) :=\bra{ \psi_{n,\vec{k}}} H\ket{\psi_{m,\vec{k}}}= \varepsilon_{m,\vec{k}}\braket{\psi_{n,\vec{k}} }{ \psi_{m,\vec{k}}}=:S_{nm}(\vec{k}) \varepsilon_{m,\vec{k}}  
$$\\
The matrix of overlaps $S_{nm}(\vec{k})$ \cite{Papaconstantopoulos_2003} can also be represented in tight-binding form and the coefficients will also vanish for large $|\vec{R}|$, because they can be computed using analogous integrals, but without the Hamiltonian. For simplicity, we will assume in the following, that the basis is orthogonal.

\subsection{Analytical Properties}\label{Analytical Properties}

The coefficients of the Hamiltonian satisfy $H^{\vec{R}}_{nm}=\overline{ H }\,\!^{-\vec{R} }_{mn}$, because $H$ is hermitian. This can be shown as follows:\\ \begin{align*}  
\sum_{\vec{R}}e^{i\vec{k}\vec{R}}H^{\vec{R}}_{nm} &=\left< \Psi_{n,\vec{k}}\middle| H \middle|\Psi_{m,\vec{k}} \right>=\overline{ \left< \Psi_{m,\vec{k}}\middle| H \middle|\Psi_{n,\vec{k}} \right> } =\sum_{\vec{R}}e^{-i\vec{k}\vec{R}}\overline{ H }\,\!^{\vec{R}}_{mn} =\sum_{\vec{R}}e^{i\vec{k}\vec{R}}\overline{ H }\,\!^{-\vec{R}}_{mn}\numberthis  
\end{align*} \\
Tight-binding models are not unique. The view of them is, that they represent the interaction of orbitals, however the orbitals are not fixed. A change of basis with a unitary matrix $U_{nm}$ on the full Hamiltonian yields another tight-binding model with the same bandstructure. In this work we normalize tight-binding models by diagonalizing $H_{nm}(\vec{k}=0)$, however this is not sufficient to make the model unique. For example with $N \geq 3$ orbitals, tight-binding models exist, which have completely different orbitals, but still the exact same bandstructure:\\
$$  
\begin{pmatrix}  
\cos(k) & 1 & 0 \\  
1 & 1 & 1 \\  
0 & 1 & -\cos(k)  
\end{pmatrix}  
\sim  
\begin{pmatrix}  
\cos(k) & 1/\sqrt{ 2 } & 1 \\  
1/\sqrt{ 2 } & 1 & 1/\sqrt{ 2 } \\  
1 & 1/\sqrt{ 2 } & -\cos(k)  
\end{pmatrix}  
$$\\
Assuming no neighbor cells $\vec{R}$ are added, these two tight-binding models can not be continuously transformed into one another without changing the bandstructure. In this case there is a discrete set of equivalent models. The optimization algorithm presented below will converge to an arbitrary solution, which makes interpolation between models from different optimization runs impossible even if they possess the same bandstructure, as, like in the example above, there might not be a continuous path from one model to the other without changing the bandstructure.

\subsection{Numerical Properties}\label{Numerical Properties}

The tight-binding formalism features $N\times N$ matrices for each $\vec{R}$, with $N$ being the number of orbitals. In numerical calculations we limit $|\vec{R}|$, such that there is a finite number of these so-called neighbor cells. We only need to store half of them, as the other half is determined by $H^{\vec{R}}_{nm}=\overline{ H }\,\!^{-\vec{R} }_{mn}$, except for $\vec{R}=0$. Assuming no more special form and $M$ stored matrices, we can compute the matrix for diagonalization for a single $\vec{k}$ in $(2M+1)N^{2}+\mathcal{O}(M)$ floating point operations (FLOPs) for real values, which results in the complexity $\mathcal{O}(MN^{2})$. The hermitian matrix diagonalization has the complexity of $\mathcal{O}(N^{3})=c N^{3}$ FLOPs. In total this leads to approximately $\left(2M+1 + cN\right)N^{2}+\mathcal{O}(M)$ FLOPs. To reduce the computational cost of the model, it is therefore essential to use as few orbitals as possible. For the neighbor count, there is more freedom, however note that $M$ grows with the distance $M\sim |\vec{R}|^{d}$ where the dimension $d$ is usually $3$.

If the bandstructure needs to be computed on a full grid of $\vec{k}$-points with $K$ points, one can use a Fast Fourier-Transform (FFT) to reduce the complexity of computing the $H_{nm}(\vec{k})$ from $\mathcal{O}(KMN^{2})$ to $\mathcal{O}(K \ln(K)^{d} N^{2})$, so the diagonalization is again the limiting factor for performance. However, this is not beneficial for the low neighbor counts, which are used in this work.

\section{Optimization technique}\label{Optimization technique}

To find tight-binding models, one has always used fitting procedures of different types. We propose a modified version of the gradient descent algorithm from machine learning. We minimize the weighted loss-function\\ \begin{align}  
 L^{2} = \frac{1}{K}\sum_{d,\alpha}w_{d,\alpha}\left|\varepsilon_{d}(H(\vec{k}_{\alpha})) - \varepsilon^{\mathrm{ref}}_{d}(\vec{k}_{\alpha})\right|^2 \label{eq:err}  
\end{align} \\
where $w_{d,\alpha}$ denotes the weights, $\varepsilon_{d}(H)$ denotes the sorted eigenvalues of the hermitian matrix $H$ and $\varepsilon_{d}^{\mathrm{ref}}$ denotes the sorted eigenvalues that we are trying to fit. The classical gradient descent algorithm with a learning rate $\eta$ can be formulated as\\ \begin{align*}  
H_{nm}(\vec{k}_{\alpha})&=:U^{\alpha}_{nd}\overline{ U^{\alpha}_{md} }\,\varepsilon_{d}(H(\vec{k}_{\alpha})) \\  
\varepsilon_{d}^{\alpha}&:=\varepsilon_{d}(H(\vec{k}_{\alpha}))\\  
H'^{\vec{R}}_{nm} &= H^{\vec{R}}_{nm} -\frac{\eta}{K}U^{\alpha}_{md}\overline{ U^{\alpha}_{nd} }w_{d,\alpha}\left(\varepsilon^{\alpha}_{d} - \varepsilon^{\mathrm{ref}}_{d}(\vec{k}_{\alpha})\right)e^{i\vec{k}_{\alpha}\vec{R}}\numberthis  
\end{align*} \\
where $\vec{k}_{\alpha}$ are the $\vec{k}$-points, at which we are fitting the band structure, $U^{\alpha}_{nm}$ are the eigenvectors of $H(\vec{k}_{\alpha})$ and $K$ is the number of $\vec{k}$-points. Here $H'^{\vec{R}}_{nm}$ are the updated parameters after a step. This algorithm works in principle, however the choice of $\eta$ is significant. The usual first acceleration technique is to use a momentum term, which isn't discussed here. The modification of this algorithm in this work is based on the observation, that the change to the bandstructure can be predicted for small steps using first order perturbation theory. In the following $H^{\vec{R}_{i}}_{nm}$ will be written as $H^{i}_{nm}$ for better legibility with Einstein sum notation.\\ \begin{align}  
d\varepsilon_{d}^{\alpha}=\overline{ U_{nd}^{\alpha} } U_{md}^{\alpha} dH_{nm}(\vec{k}_{\alpha})=\overline{ U_{nd}^{\alpha} } U_{md}^{\alpha} e^{i\vec{k}_{\alpha}\vec{R}_{i}} dH^{i}_{nm}  
\end{align} \\
It appears, that for bandstructures this prediction also holds for rather large steps. This would e.g.~not be true for the classical deep neural network. Thus, it leads naturally to the idea of finding a least-squares solution to the step $dH_{nm}^{i}$ given the necessary change $d\varepsilon_{d}^{\alpha}=\varepsilon^{\alpha}_{d} - \varepsilon^{\mathrm{ref}}_{d}(\vec{k}_{\alpha})$. The least-squares solution can be formulated in the following way:\\ \begin{align*}  
d\varepsilon_{d}^{\alpha}&= A_{(\alpha d)(inm)} dH^{i}_{nm} \\  
A_{(\alpha d)(inm)}&:=\overline{ U_{nd}^{\alpha} } U_{md}^{\alpha} e^{i\vec{k}_{\alpha}\vec{R}_{i}} \numberthis\\  
\tilde{L}^{2}&=w_{d,\alpha}\left| d\varepsilon_{d}^{\alpha} - (\varepsilon^{\mathrm{ref}}_{d}(\vec{k}_{\alpha})-\varepsilon_{d}^{\alpha}) \right|^{2} \\  
\frac{1}{2}\frac{\partial \tilde{L}^{2}}{\partial dH^{i}_{nm}}&=w_{d,\alpha}( d\varepsilon_{d}^{\alpha}  - (\varepsilon^{\mathrm{ref}}_{d}(\vec{k}_{\alpha})-\varepsilon_{d}^{\alpha}))\overline{ A }_{(\alpha d)(inm)} \mathop{=}\limits^{!} 0\\  
\Rightarrow\quad  dH^{j}_{kl} A_{(\alpha d)(jkl)} \overline{ A }_{(\alpha d)(inm)}w_{d,\alpha} &= w_{d,\alpha}(\varepsilon^{\mathrm{ref}}_{d}(\vec{k}_{\alpha})-\varepsilon_{d}^{\alpha})\overline{ A }_{(\alpha d)(inm)} =: v_{inm}\numberthis\\  
F_{(jkl)(inm)} &:= \overline{ A }_{(\alpha d)(jkl)} A_{(\alpha d)(inm)} w_{d,\alpha}  
\end{align*} \\
This is a system of linear equations of the form $F\vec{dH}=\vec{v}$, which needs to be solved to find the step in the coefficients $dH_{kl}^{i}$. The matrix $F$ requires $(MN^{2})^{2}$ elements, which is too large to store or invert directly. To simplify this, the action of $F$ can be written as\\ \begin{align}  
F_{(jkl)(inm)}dH^{i}_{nm}=\sum_{\alpha}e^{-i\vec{k}_{\alpha}\vec{R}_{j}} \sum_{d} w_{d,\alpha} U_{kd}^{\alpha} \overline{ U_{ld}^{\alpha} } \sum_{nm}\overline{ U_{nd}^{\alpha} } U_{md}^{\alpha}  \sum_{i}e^{i\vec{k}_{\alpha}\vec{R}_{i}}dH^{i}_{nm} \label{eq:ls-lin}  
\end{align} \\
With this order of summation the action of $F$ can be computed in $K(4MN^{2}+6N^{3})$ FLOPs, where $K$ is the number of $\vec{k}$-points. This is usually the most efficient way to compute the sum. It can easily be parallelized over $K$. The $\sqrt{ w_{d,\alpha} }U^{\alpha}_{kd}\overline{ U^{\alpha}_{ld} }$ can be precomputed to make it slightly faster within reasonable memory bounds of $KN^{3}$ values. For comparison, the diagonalization takes $K(2MN^{2}+cN^{3})$ FLOPs, where $c$ is a constant depending on the diagonalization procedure. To solve the system of linear equations (\ref{eq:ls-lin}), we use the conjugate gradient (CG) algorithm \cite{cg} with a limited number of iteration steps. Due to the step limitation, the CG has the same time scaling as the diagonalization. In practice, we have found that 6 steps of the CG are usually enough to ensure good convergence of the full procedure.

The conjugate gradient method allows the use of a preconditioner $L$.\\ \begin{align}  
F\vec{x}=\vec{b}\quad  \Leftrightarrow\quad L^{\dagger}FL\vec{y}=L^{\dagger}\vec{b},\quad \vec{x}=L\vec{y},\quad L \in GL(n,\mathbb{C})  
\end{align} \\
We choose $L$ from the Cholesky decomposition $\left( \sum_{\alpha}e^{i\vec{k}_{\alpha}(\vec{R}_{i}-\vec{R}_{j})} \right)_{ij}^{ + }=LL^{\dagger}$. The idea behind this choice is, that in case of $U^{\alpha}_{nm}=\delta_{nm}$ and in case the above matrix $LL^{\dagger}$ is invertible, we get $L^{\dagger}FL=\mathds{1}$. This preconditioner is best used, when the $\vec{k}$-points are not on a regular grid, as it becomes trivial for regular grids.

\begin{figure}[ht]
\centering
\includegraphics[width=\textwidth,keepaspectratio]{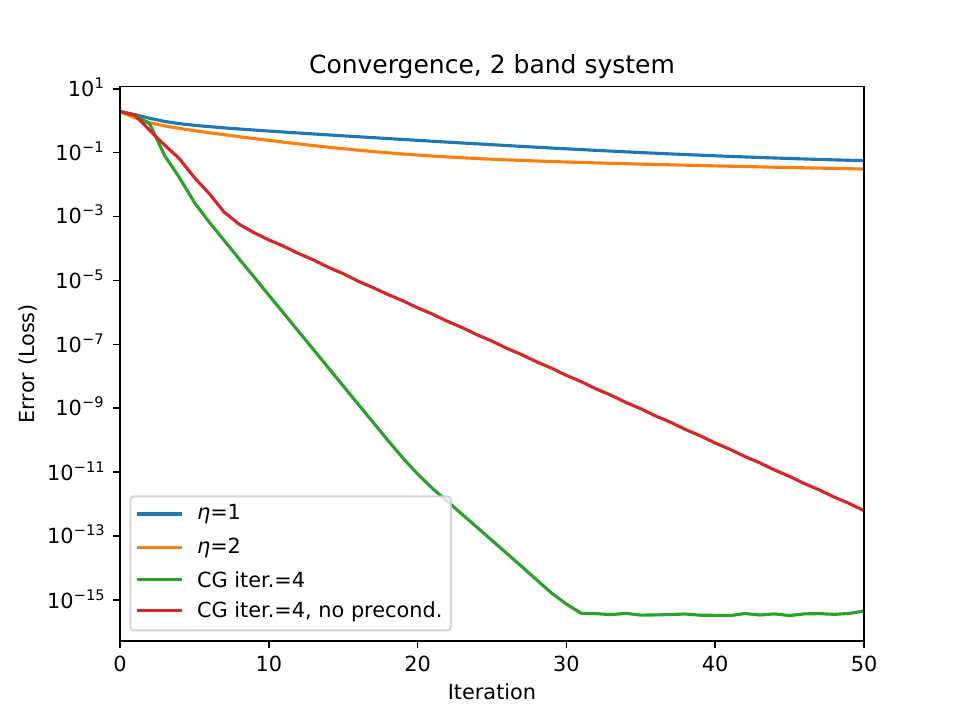}
\caption{Convergence of Gradient Descent and the improved method based on least
squares. The iterations of the x-axis are those of the whole procedure.
The CG iterations have been limited to 4.
}\label{1fig1}
\end{figure}

The numerical implementation of gradient descent, non-preconditioned CG and preconditioned CG are compared in figure \ref{1fig1} for a one dimensional system with two bands and one neighbor term. The reference band values $\varepsilon(k)$ have been sampled from an exact solution in an interval $k\in[ -\pi/2, \pi]$ which doesn't cover the full $k$-space $[ -\pi, \pi]$. The reference model is the following model, which has a band crossing:\\
$$  
H(k)=\begin{pmatrix}  
2\cos(k) & 0 \\  
0 & 1-2\cos(k)  
\end{pmatrix}  
$$

In the general case the procedure is the following:

\begin{enumerate}
\def\labelenumi{\arabic{enumi}.}
\item
  Start with flat bands with energies matching the reference at the $\Gamma$-point. To eliminate starting subspace issues, randomize all parameters by a small amount.
\item
  Fit the bandstructure with an increasing amount of $\vec{k}$-points, expanding from the $\Gamma$-point. In our fits, the data has been added in 10 equal batches. Each time a new batch is added, one iteration of least-squares fitting is applied.
\item
  Repeat step 1, 2 several times and use the best fitting model to start the next step. In this work 10 restarts have been used unless explicitly stated otherwise, but in practice 3 would have been sufficient to save some computation time.
\item
  Fit the model with the least-squares procedure described in this paper. Each time the procedure converges, a small randomization scaled to the current loss is applied, and the fit is continued. We repeated this 10 times. Our convergence criterion was a change of loss smaller than $10^{-3}$ per iteration. One can get near optimal results (within $\sim 20\%$) much faster by skipping the randomized restarts. It is an open question why they help in the first place.
\item
  Symmetrically add more neighbors to the model by increasing the maximal radius $\vec{R}$ with $\vec{R}$ in unscaled real space. Then repeat step 4.
\end{enumerate}

We started the procedure with 3 symmetrically equivalent groups of neighbors. For the fitted data we have selected a range of bands and weights by hand, but this could be easily automated by using an energy cutoff. The initialization steps have been chosen to avoid initializing in a bad local minimum with spurious band crossings. The reason for adding neighbors incrementally is to avoid overfitting. To finalize the fit, one can optionally do one more step using $M=K$ neighbors to eliminate all errors on the reference data. This turns the procedure into an interpolation scheme instead of a fitting scheme.

The procedure suggested here is a deterministic minimization. We want to point out that it contains two measures in order to avoid getting stuck in local minima: in step 2 we are successively extending the training data set by increasing the number of $\vec{k}$-points used in the optimization expanding from the Gamma point. This means that the loss function is changing during the procedure, which avoids getting stuck in local minima. We are starting with the data points closest to the $\Gamma$-point and adding more data points further away. In this way the Taylor-expansion at the $\Gamma$-point is matched early, solving band-sorting issues. Secondly, in step 3 randomized restarts are used in the beginning, where the best among 10 random models is chosen without fully optimizing them. This allows to find a good starting point for the full optimization in steps 4 and 5.

In tight-binding calculations the on-site and nearest neighbor hopping parameters are usually of particular importance. Here, the on-site parameters are the $H^{\vec{R}}_{nm}$ parameters for $\vec{R}=0$ and the nearest neighbor parameters the ones for the smallest non-zero $|\vec{R}|$. The on-site parameters are initialized in step 1 to match the energies at the $\Gamma$-point, while the other hopping parameters are set to almost zero with a small random perturbation. In step 2 only a fixed subset of parameters $H^{\vec{R}}_{nm}$ up to some $|\vec{R}|\leq R_{\mathrm{ max }}$ are optimized. In step 5, the number of neighbors is increased gradually. Note, that the on-site and nearest neighbor parameters are determined only at the very end of the procedure, as they are allowed to change in each optimization step.

In eq. (\ref{eq:tb}) we focused on tight-binding models for a single atomic structure. However, the procedure suggested here can also be generalized to create so-called transferable tight-binding models, which are valid for a range of different structures. This generalization is described in the appendix.

\section{Application to selected materials}\label{Application to selected materials}

\begin{figure}[ht]
\centering
\includegraphics[width=\textwidth,keepaspectratio]{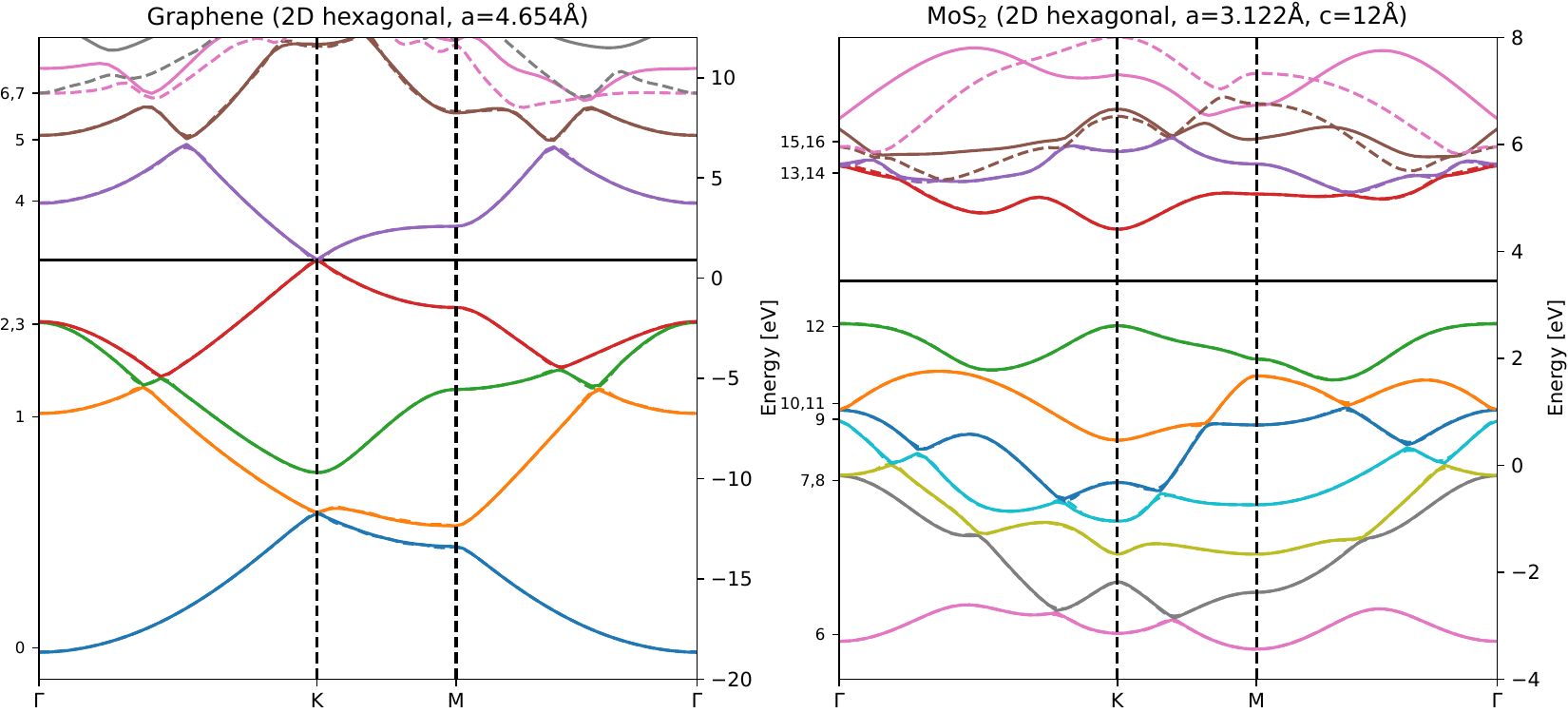}
\caption{Graphene and $\mathrm{ MoS}_{2}$ with neighbors $(1,0,0)$,
$(3/2,\sqrt{ 3 }/2,0)$. The reference DFT data is dashed. The top
two bands are fitted with a low weight.
}\label{1fig2}
\end{figure}

\begin{figure}[ht]
\centering
\includegraphics[width=\textwidth,keepaspectratio]{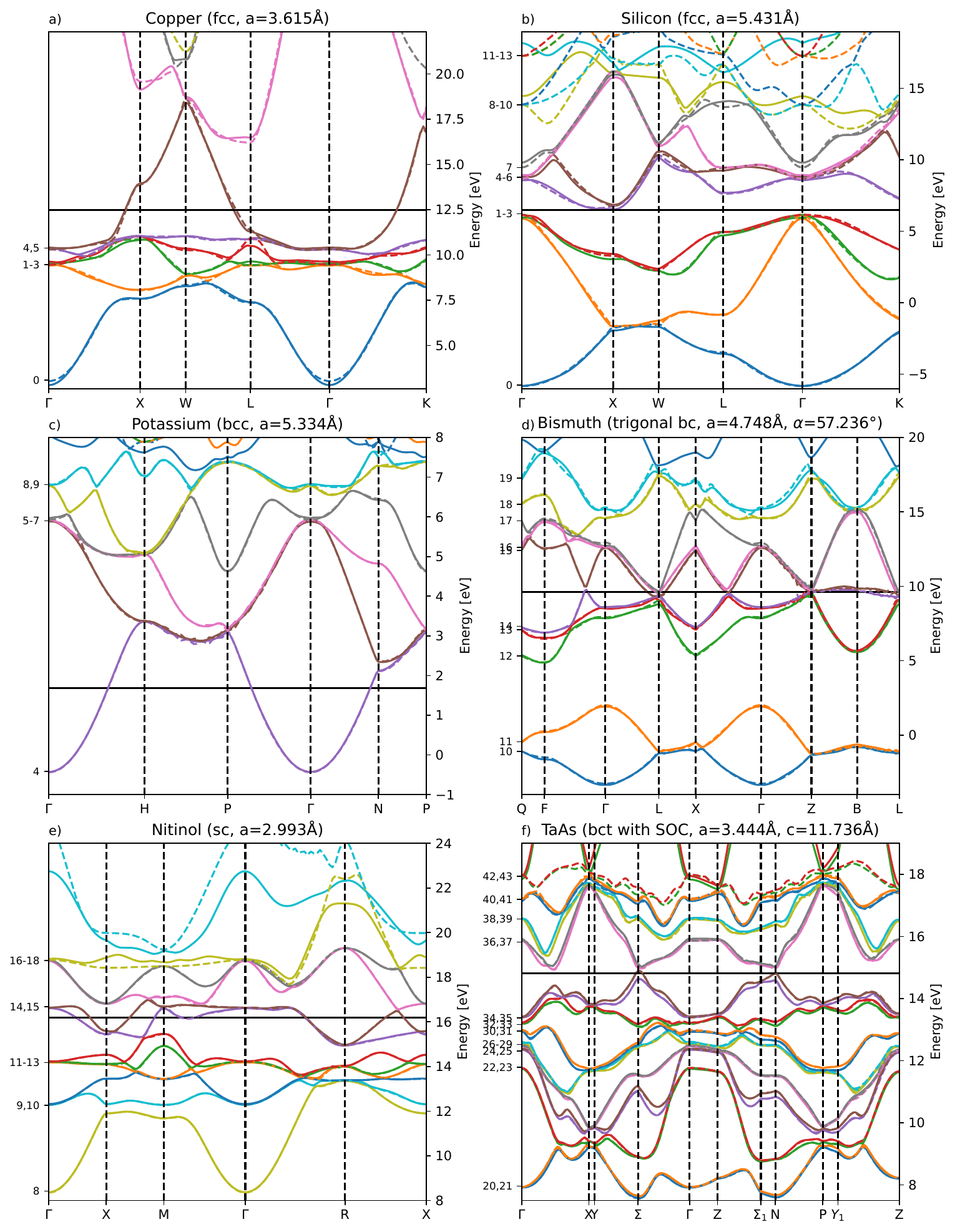}
\caption{Bandstructures for Cu and Si with neighbors $(0,1/2,1/2)$,
$(0,0,1)$, $(1/2,1/2,1)$, K with neighbors $(1/2,1/2,1/2)$,
$(0,0,1)$, $(0,1,1)$, Bi with neighbors $(0,0.958,0)$,
$(0.553,0,0.833)$, $(1.106,0,-0.833)$, NiTi with neighbors
$(0,0,1)$, $(0,1,1)$, $(1,1,1)$, $(0,0,2)$ and TaAs with cubic
symmetrized normalized neighbors $(1/2,1/2,1/2)$, $(0,0,1)$,
$(1/2,1/2,3/2)$. The reference DFT data is dashed.
}\label{1fig3}
\end{figure}

We have tested the procedure for various materials with different crystal structures. It has shown to work reliably for all of them to produce good fitting tight-binding models with a low neighbor count. For all the 3D structures we have computed a $24^{3}$ lattice of $\vec{k}$-points using the DFT package \emph{Quantum Espresso} \cite{QE-2017,QE-2009,doi:10.1063/5.0005082}. Only for TaAs we have enabled spin orbit coupling (SOC). Bismuth would in principle also need SOC for realistic results. We have used KJPAW pseudo-potentials for the PBE GGA-functional from \emph{PSLibrary} \cite{pslibrary}. For the self-consistent computation of the electron density, we used a lattice of $10^{3}$ $\vec{k}$-points. We also calculated graphene and $\mathrm{ MoS }_{2}$ as 2D examples. For both 2d examples we used RRKJUS pseudo-potentials, a $9^{2}$ grid for the self-consistency calculation and a $24^{2}$ grid for the fit. For all fits, we have selected a range of bands by hand and then weighted the top two bands with $0.01$ in the loss function. The results are shown in figure \ref{1fig2} and figure \ref{1fig3}.

\section{Wavefunctions and Orbitals}\label{Wavefunctions and Orbitals}

In other tight-binding procedures, one often gets the orbitals explicitly. \cite{TBPao,Wannier90} The procedure presented here does not provide the orbitals, as the tight-binding model does not contain enough information. There is a relation between the Bloch-eigenfunctions $\Psi$ from DFT and the Orbitals $\phi$. The eigenspaces spanned by the tight-binding eigenfunctions and the known eigenfunctions $\Psi$ need to be equal. If the eigenvalues are all unique, the known eigenfunctions equal the tight-binding eigenfunctions up to a phase. They can then be transformed back to the tight-binding basis functions $\psi$ using the unitary matrix $U_{nm}(\vec{k})$, which diagonalizes the tight-binding model. The freedom of choice of $\Psi$ can equivalently be seen as a freedom of choice of $U_{nm}(\vec{k})$.\\
$$  
\Psi_{n,\vec{k}}(\vec{r})=\overline{ U_{mn} }(\vec{k})\psi_{m,\vec{k}}(\vec{r})  
$$\\
There are two problems with this approach. First, only $\vec{k}$-points with unique eigenvalues can be used. Second, the phases need to be chosen in a way that minimizes oscillations as discussed in \cite{MazariReview} for the application in \emph{Wannier90}. Then one can compute the orbitals from the integral\\
$$  
\phi_{n}(\vec{r})=\frac{1}{V_{BZ}}\int_{1.BZ} \psi_{n,\vec{k}}(\vec{r}) d\vec{k}  
$$\\
Here one needs to use correct integration weights to account for the fact, that highly symmetric $\vec{k}$-points with non-unique eigenvalues are skipped.

A further procedure similar to \emph{wannier90} \cite{MazariReview} is needed to find the best choice of these phases to construct well localized orbitals for given tight-binding models. It has not been implemented in this work.

\section{Comparison with Wannier90}\label{Comparison with Wannier90}

\begin{figure}[ht]
\centering
\includegraphics[width=\textwidth,keepaspectratio]{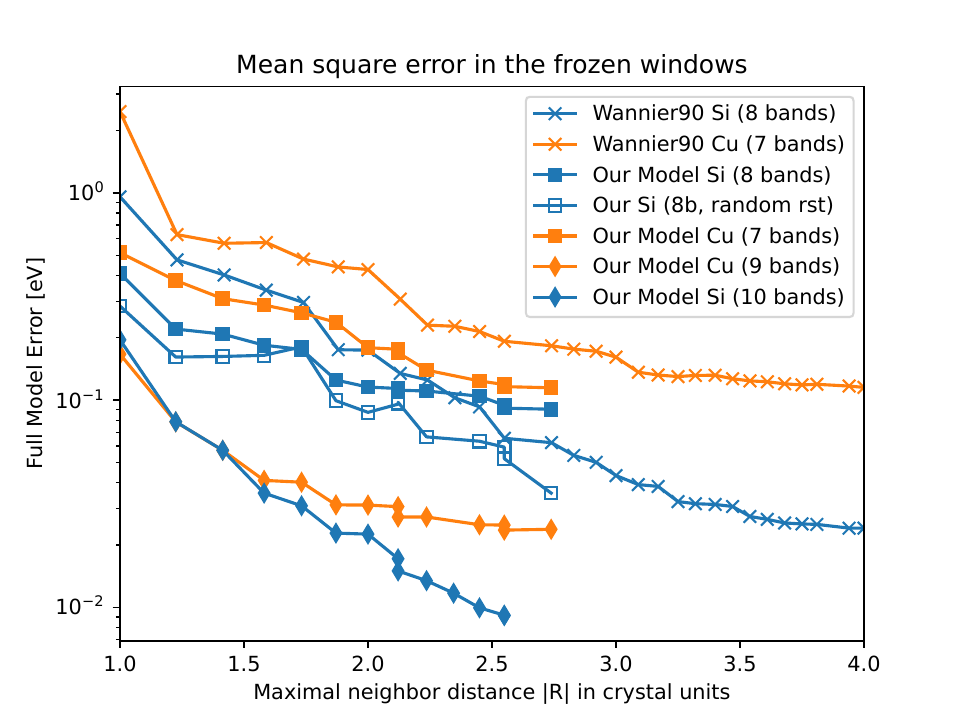}
\caption{Error of Wannier90 ($\times$) compared to our method ($\square$),
when used for interpolation with a finite number of neighbors up to some
maximal magnitude. The open squares show the result with randomized
restarts. The diamonds are results with increased band count.
}\label{1fig5}
\end{figure}

We have chosen to compare our method with the established code \emph{wannier90} \cite{Wannier90}. Both methods produce asymmetric tight-binding models, however \emph{wannier90} also produces the momentum matrix elements. We have found that due to the disentanglement step in \emph{wannier90} there is a systematic error for metals, which is not present in our method. For comparison, models for the reference materials silicon and copper have been created both using wannier90 and using our method. The data for these has been computed as in section \ref{Application to selected materials}, but with a $8^{3}$ grid for self-consistency and a $16^{3}$ grid without symmetry for the raw data. Both codes don't use any symmetry. The frozen window in wannier90 has been chosen to cover the bands, which have weight $1$ in our fit procedure. Our fits have been computed without the proposed additional randomization step, however some fits with the randomization steps are also included in figure \ref{1fig5} for reference. In \emph{wannier90} it is necessary to choose starting orbitals by hand. For silicon, we used one $s$- and 3 $p$-orbitals each at the two inequivalent atomic positions. For copper, we used 5 $d$-orbitals at the copper site and two $s$-orbitals positioned at the copper site and an interatomic site, respectively.

\begin{figure}[ht]
\centering
\includegraphics[width=\textwidth,keepaspectratio]{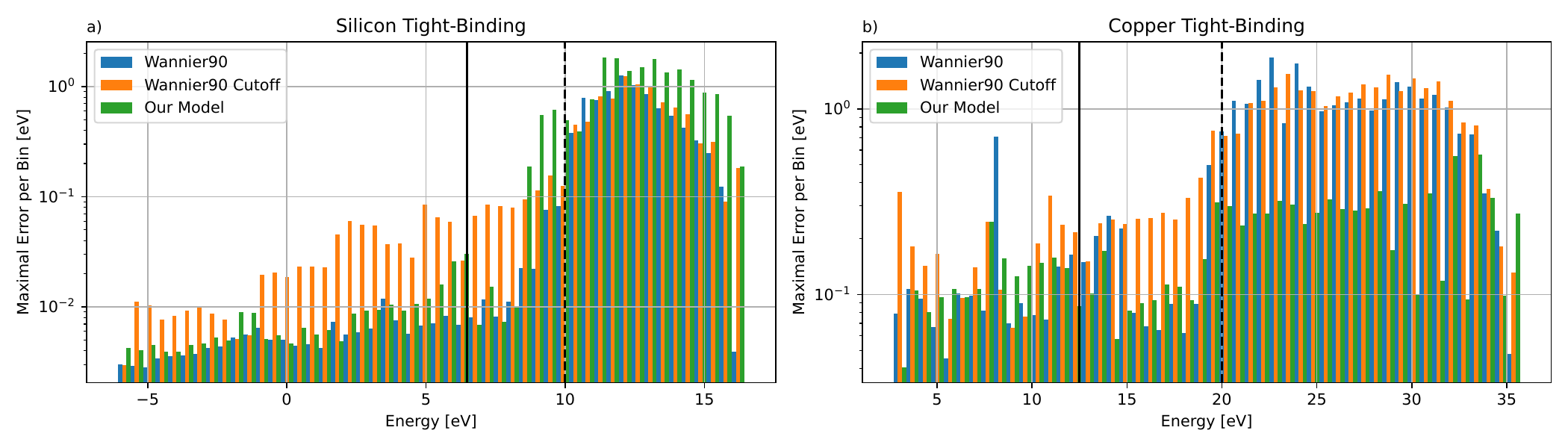}
\caption{Maximal errors of the tight-binding models for wannier90 with all
neighbors, our model best model with 16 neighbors and cut-off
\emph{Wannier90} to match our model in neighbor count. All models are
fit on a $16^{3}$ grid, but evaluated on a $24^{3}$ grid. The
vertical line is the Fermi energy and the dashed vertical line is the
upper edge of the frozen energy window.
}\label{1fig4}
\end{figure}

While the training data set consists of a $16^{3}$ $\vec{k}$-point grid, we also created a test data set using an additional DFT calculation on a $24^{3}$ grid to avoid any overfitting bias. Figure \ref{1fig4} shows the results of \emph{wannier90} and our method for 16 neighbors by plotting the maximal error of the model over the energy range as a kind of histogram. The values are calculated with respect to the test data set. At this neighbor count, the full \emph{wannier90} model has still more precision, however only in some regions and with much more parameters. It also has a high error at $8\mathrm{ eV }$ in copper, which is indicative of overfitting. It disappears in the cutoff model, as the cutoff smoothes the function.

To compare our model to wannier90 it would be unfair to use the loss from (\ref{eq:err}), as that is what the fit procedure is minimizing, while wannier90 doesn't know about our choice of band weights. Instead, the error metric is chosen to match wannier90's frozen energy window, by only including data points in (\ref{eq:err}), which are in the frozen window. To make it even more fair to wannier90, the bands are not assumed to be sorted and bands can be skipped in some places. This is done by finding the minimal perfect matching of the squared errors. With this definition of the error, figure \ref{1fig5} shows how wannier90 models compare to our models. The models, which are all fitted on the $16^{3}$ grid, are again evaluated on the $24^{3}$ grid to avoid any overfitting bias. A cutoff has been used on the wannier90 model, such that terms with $|\vec{R}|>R$ are removed. Our models for different maximal neighbor distances come from saving checkpoints during the fit procedure, where the neighbors are successively added. Note, that the fit procedure will converge against an error of 0 on all bands when given $M \geq K$ neighbors, which turns it into an interpolation method, however on the test dataset with a finer grid it will still have a finite error.

\begin{figure}[ht]
\centering
\includegraphics[width=\textwidth,keepaspectratio]{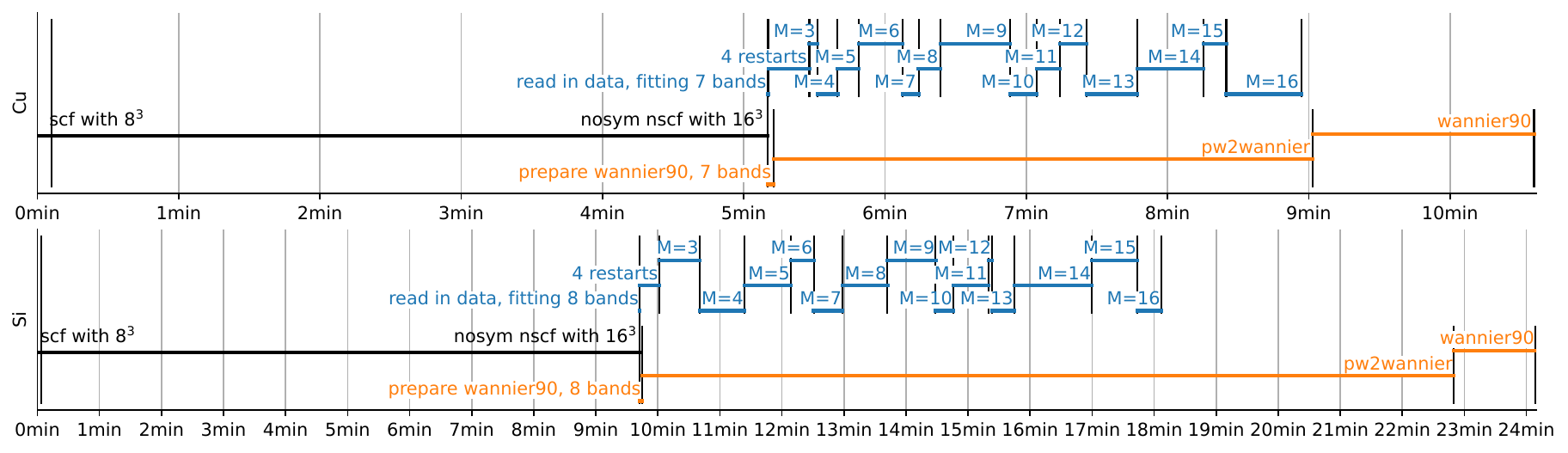}
\caption{The timing data from the full procedure for wannier90 and our method (no
random restarts) with equal band counts. The non-symmetric DFT
calculation \emph{nscf}, which is necessary for \emph{wannier90} and the
\emph{pw2wannier90} step, which computes the projections and overlaps
for \emph{wannier90}, takes most of the time.
}\label{1fig6}
\end{figure}

The time of all steps has been measured on a PC with an Intel Core i7-8700 (12 logical cores, 4.6GHz). The time for the fit of silicon with the described randomized restarts in figure \ref{1fig5} took 38 minutes in total. The time of the fitting procedure scales with $\mathcal{O}(KN^{3}+KMN^{2})$ where $N$ is the number of orbitals and $M$ is the number of neighbors. For a chosen cutoff $R$ in $d$ dimensions, this results in a scaling of $\mathcal{O}(KN^{3}+KN^{2}R^{d})$ per iteration, however the first term usually has the larger constant, as can be seen in figure \ref{1fig6}, because the time for each $M$ is roughly equal. The total time scales with $\mathcal{O}(KN^{3}R^{d})$. The space complexity is $\mathcal{O}(MN^{2}+KN)$ or $\mathcal{O}(R^{d}N^{2}+KN)$. The wannier90 procedure starts with the program \emph{pw2wannier} to compute the overlap-matrices $A^{(k)}_{nm}$ and $M^{(k,b)}_{nm}$ \cite{SouzaMLWF} from the DFT data and then the program \emph{wannier90} to disentangle and compute the maximally localized Wannier-functions. The time complexity for \emph{pw2wannier} is dependent on the energy cutoff, which has been used in the non-self consistent DFT calculation (\emph{nscf}). The energy cutoff determines the cutoff of the plane wave basis. In the following $r_{G}$ denotes the radius of this cutoff for the plane wave reciprocal lattice vectors $\vec{G}$. Then we can write the time complexity as $\mathcal{O}(r_{G}^{d} KN_{r}(N+2dN_{r}))$ where $K$ is the number of $\vec{k}$-points used in the non-symmetric \emph{nscf} calculation, $N$ is the number of Wannier-bands and $N_{r}$ is the number of non-excluded bands that have been computed in the \emph{nscf} calculation. The space complexity is $\mathcal{O}(KN_{r}(N+2dN_{r}))$. The number of computed bands $N_{r}\geq N$ is usually chosen to be a few bands more than the used projections and then disentanglement is used. In the test $N_{r}$ has been chosen as $N_{r}=N+2$. Then \emph{wannier90} is run. Without disentanglement, it has the time complexity of $\mathcal{O}(KdN^{3})$ for initialization and $\mathcal{O}(KN^{3})$ per \emph{wannierization} iteration \cite{MazariMLWF}. The space complexity is $\mathcal{O}(KdN^{2})$ due to the need to compute the $M^{(k,b)}_{nm}$ matrix once.

\section{Fermi Surface}\label{Fermi Surface}

\begin{figure}[ht]
\centering
\includegraphics[width=\textwidth,keepaspectratio]{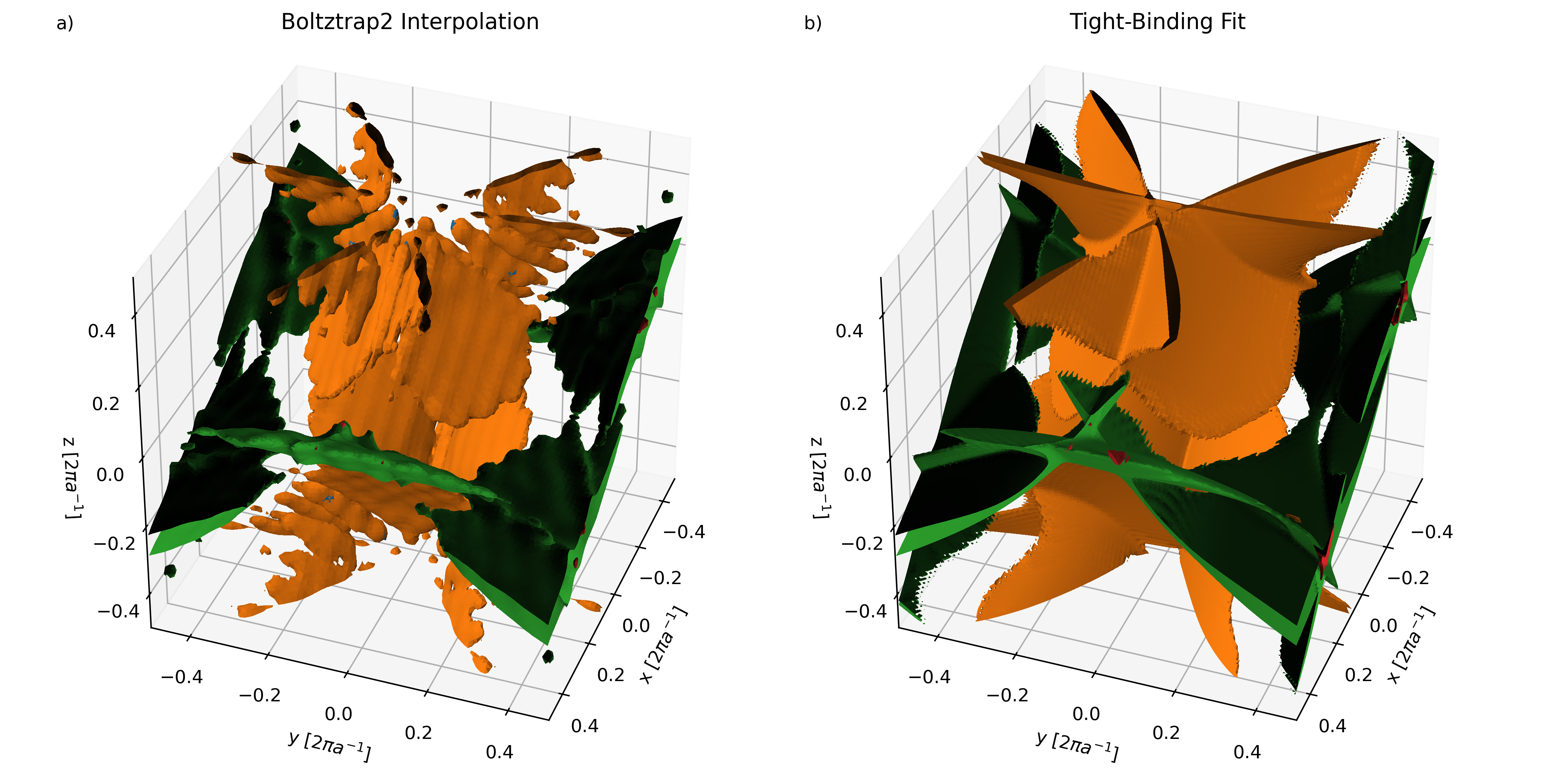}
\caption{Fermi-surface plots for bismuth using Boltztrap interpolation (left) and
a tight-binding fit (right)
}\label{1fig7}
\end{figure}

Since our tight-binding models work with very few neighbors, they produce smooth results for the Fermi-surface. In particular when comparing to roughness minimizing FFT interpolation \cite{Pickett1988SmoothFI} as implemented in Boltztrap2 \cite{boltztrap2}, our method yields much smoother results, which is ideal for computing group velocities. As seen in figure \ref{1fig7}, Boltztrap2 fails to faithfully produce Fermi-surfaces suitable for integration of transport properties depending on derivatives, when there are fine details, as with semimetals. As visible in figure \ref{1fig4}b at $8\mathrm{ eV }$, wannier90 can produce some sharp features as well, which are not contained in the data, because it uses many neighbor terms. However, generally those models behave similar to the models, which are produced in this work, as localized orbitals reduce far neighbor terms, too.

\section{Conclusion}\label{Conclusion}

In this work we devised a procedure to construct tight-binding models for the electronic structure of materials by training on DFT band structures. The method employs least-squares minimization based on first order perturbation theory using conjugate gradients. Special features are a preconditioner, gradual data introduction from the $\Gamma$-point, and stochastic refinement by randomized restarts.

In contrast to conventional gradient decent methods, this method features much faster convergence. It also allows systematic control of accuracy, as more neighbors can be added successively until a desired accuracy is reached. We have demonstrated that a high accuracy can be reached already with few neighbors. The resulting tight-binding models possess smooth band structures, which is useful for calculation and interpolation\\
of group velocities and band masses, as well as determination of Fermi surfaces. The models represent the band structure data with a minimum amount of parameters. Orbitals are chosen automatically without the need of human control.\\

\section{Data availability statement}\label{Data availability statement}

The code of the python package developed and used in this work is available on GitHub at \url{https://github.com/redweasel/tight-binding.} The code to produce the plots from this paper is published at \cite{epubdata}.\\

\section{Acknowledgments}\label{Acknowledgments}

We would like to thank Adrian Braun and Malte Rösner for helpful discussions. We also thank the HPC.NRW Team for their support, as some of the DFT calculations have been done on the Bielefeld GPU Cluster. Financial support from the DFG via the research group FOR2692, grant number~397171440 is gratefully acknowledged.

\appendix

\section{General Optimization Method}\label{General Optimization Method}

In this manuscript we have focused so far on simple tight-binding models of the form\\
$$  
  H_{nm}(\vec{k}_\alpha) = \sum_{i} \lambda_i(\vec{k}_\alpha) H^i_{nm}  
$$\\
where the functions $\lambda_i(\vec{k})=e^{i\vec{k}\vec{R}_i }$ are plane waves. However, the procedure\\
presented here also works for other sets of $\vec{k}$-dependent functions. Moreover, one can even include further parameters on which the tight-binding parameters may depend. This is useful for so-called transferable tight-binding models, in which the tight-binding parameters may depend on the atomic positions. In such cases the functions $\lambda_i$ will depend on $\vec{k}$ and a collection of further parameters $\vec{p}$ such that $\lambda_i(\vec{k}_\alpha,\vec{p}_\beta)$. To generate such a transferable tight-binding model one would want to train it on a large set of DFT calculations covering different atomic structures of interest. If we number all parameter combinations $(\vec{k},\vec{p})$ in the training data set by a single counter $\alpha$ and collect the set of functions $\lambda_i$ by a vector $\vec{\lambda}$, the numerical method used here extends to a more general optimization task:\\ \begin{align*}    
H_{nm}(\vec{\lambda})&=H_{nm}^{i}\lambda_{i} = U_{nd}(\vec{\lambda})\varepsilon_{d}(\vec{\lambda})\overline{ U_{md} }(\vec{\lambda}) \\  
&\mathrm{ min }_{H_{nm}^{i}}(|w_{d\alpha}(\varepsilon_{d}(\vec{\lambda}^{(\alpha)}) - \varepsilon_{d}^{\mathrm{ ref }}(\vec{\lambda}^{(\alpha)}) )|)\numberthis  
\end{align*} \\
The algorithm can then be formulated as follows:\\ \begin{align*}  
d\varepsilon_{d}^{\alpha}&=\overline{ U_{nd}^{\alpha} } U_{md}^{\alpha} dH_{nm}(\vec{\lambda}^{(\alpha)})=\overline{ U_{nd}^{\alpha} } U_{md}^{\alpha} \lambda^{(\alpha)}_{i} dH^{i}_{nm}=: A_{(\alpha d)(inm)} dH^{i}_{nm} \\  
&\mathrm{ min }_{\Delta H_{nm}^{i}}(|w_{d\alpha}( A_{(\alpha d)(inm)}\Delta H^{i}_{nm} - \Delta\varepsilon_{d}^{\alpha} )|)\numberthis  
\end{align*} \\
In the main text we used $\lambda^{(\alpha)}_{i}=e^{i\vec{k}_{\alpha}\vec{R}_{i}}$, but one could also add some other ansatz like $\lambda_{ij}^{(\alpha)}=(A_{\alpha}^{-1})_{ij}$ for changes of the unit cell, where $A$ is the lattice matrix, or $\lambda_{ij}^{(\alpha)}=e^{i\vec{r}_{i}^{(\alpha)} \vec{G}_{j}}$ where $\vec{r}_{i}^{(\alpha)}$ are the atom positions and $\vec{G}_j$ are the reciprocal lattice vectors. One $\lambda$ should always be constant $\lambda_{0}=1$. For simple tight-binding models this is the $\lambda$ for $\vec{R}=0$. For interpolation, the $\lambda_i$ should always be chosen to be continuous functions of the data. The preconditioner can also be formulated generally as\\
$$    
\left(\overline{ \lambda^{(\alpha)}_{i} } \lambda^{(\alpha)}_{j} \right)_{ij}^{ + }=LL^{\dagger}  
$$\\
such that for $U(\vec{\lambda})=\mathds{1}$ the inverted matrix $L^{\dagger}FL$ becomes trivial if the matrix was invertible. The matrix is invertible if the data points $\alpha$ span the whole configuration space, otherwise overfitting will necessarily occur.\\
$$  
(LL^{\dagger}FL)_{in}=\left(\overline{ \lambda^{(\alpha)}_{i} } \lambda^{(\alpha)}_{j}\right)_{ij}^{ + } \overline{ \lambda^{(\alpha)}_{j} } \lambda^{(\alpha)}_{k} L_{kn}\mathop{=}\limits^{\mathrm{ full\ rank }}L_{in}  
$$

\section*{References}

\bibliographystyle{iopart-num}
\bibliography{refs}

@article{SlaterKoster,
  title = {Simplified LCAO Method for the Periodic Potential Problem},
  author = {Slater, J. C. and Koster, G. F.},
  journal = {Phys. Rev.},
  volume = {94},
  issue = {6},
  pages = {1498--1524},
  numpages = {0},
  year = {1954},
  month = {Jun},
  publisher = {American Physical Society},
  doi = {10.1103/PhysRev.94.1498},
  url = {https://doi.org/10.1103/PhysRev.94.1498}
}

@article{Goringe_1997,
doi = {10.1088/0034-4885/60/12/001},
url = {https://dx.doi.org/10.1088/0034-4885/60/12/001},
year = {1997},
month = {dec},
publisher = {},
volume = {60},
number = {12},
pages = {1447},
author = {C M Goringe and D R Bowler and E Hernández},
title = {Tight-binding modelling of materials},
journal = {Reports on Progress in Physics}
}

@article{Papaconstantopoulos_2003,
doi = {10.1088/0953-8984/15/10/201},
url = {https://dx.doi.org/10.1088/0953-8984/15/10/201},
year = {2003},
month = {mar},
publisher = {},
volume = {15},
number = {10},
pages = {R413},
author = {D A Papaconstantopoulos and M J Mehl},
title = {The Slater–Koster tight-binding method: a computationally efficient and accurate
approach},
journal = {Journal of Physics: Condensed Matter}
}

@book{Papaconstantopoulos_book,
	author        = "D. A. Papaconstantopoulos",
	title         = "{Handbook of the Band Structure of Elemental Solids}",
	publisher     = "Springer",
	address       = "New York, NY",
	year          = "2015",
	url           = "https://doi.org/10.1007/978-1-4419-8264-3",
}

@article{AndersenNMTO,
  title = {Muffin-tin orbitals of arbitrary order},
  author = {Andersen, O. K. and Saha-Dasgupta, T.},
  journal = {Phys. Rev. B},
  volume = {62},
  issue = {24},
  pages = {R16219--R16222},
  numpages = {0},
  year = {2000},
  month = {Dec},
  publisher = {American Physical Society},
  doi = {10.1103/PhysRevB.62.R16219},
  url = {https://doi.org/10.1103/PhysRevB.62.R16219}
}

@article{MazariMLWF,
  title = {Maximally localized generalized Wannier functions for composite energy bands},
  author = {Marzari, Nicola and Vanderbilt, David},
  journal = {Phys. Rev. B},
  volume = {56},
  issue = {20},
  pages = {12847--12865},
  numpages = {0},
  year = {1997},
  month = {Nov},
  publisher = {American Physical Society},
  doi = {10.1103/PhysRevB.56.12847},
  url = {https://doi.org/10.1103/PhysRevB.56.12847}
}

@article{SouzaMLWF,
  title = {Maximally localized Wannier functions for entangled energy bands},
  author = {Souza, Ivo and Marzari, Nicola and Vanderbilt, David},
  journal = {Phys. Rev. B},
  volume = {65},
  issue = {3},
  pages = {035109},
  numpages = {13},
  year = {2001},
  month = {Dec},
  publisher = {American Physical Society},
  doi = {10.1103/PhysRevB.65.035109},
  url = {https://doi.org/10.1103/PhysRevB.65.035109}
}

@article{MazariReview,
  title = {Maximally localized Wannier functions: Theory and applications},
  author = {Marzari, Nicola and Mostofi, Arash A. and Yates, Jonathan R. and Souza, Ivo and Vanderbilt, David},
  journal = {Rev. Mod. Phys.},
  volume = {84},
  issue = {4},
  pages = {1419--1475},
  numpages = {0},
  year = {2012},
  month = {Oct},
  publisher = {American Physical Society},
  doi = {10.1103/RevModPhys.84.1419},
  url = {https://doi.org/10.1103/RevModPhys.84.1419}
}

@article{Wannier90,
title = {wannier90: A tool for obtaining maximally-localised Wannier functions},
journal = {Computer Physics Communications},
volume = {178},
number = {9},
pages = {685-699},
year = {2008},
doi = {10.1016/j.cpc.2007.11.016},
url = {https://doi.org/10.1016/j.cpc.2007.11.016},
author = {Arash A. Mostofi and Jonathan R. Yates and Young-Su Lee and Ivo Souza and David Vanderbilt and Nicola Marzari}
}

@article{TBPao,
  title = {Accurate tight-binding Hamiltonian matrices from ab initio calculations: Minimal basis sets},
  author = {Agapito, Luis A. and Ismail-Beigi, Sohrab and Curtarolo, Stefano and Fornari, Marco and Nardelli, Marco Buongiorno},
  journal = {Phys. Rev. B},
  volume = {93},
  issue = {3},
  pages = {035104},
  numpages = {9},
  year = {2016},
  month = {Jan},
  publisher = {American Physical Society},
  doi = {10.1103/PhysRevB.93.035104},
  url = {https://doi.org/10.1103/PhysRevB.93.035104}
}

@article{Gresch2018,
  title = {Automated construction of symmetrized Wannier-like tight-binding models from ab initio calculations},
  author = {Gresch, Dominik and Wu, QuanSheng and Winkler, Georg W. and H\"auselmann, Rico and Troyer, Matthias and Soluyanov, Alexey A.},
  journal = {Phys. Rev. Mater.},
  volume = {2},
  issue = {10},
  pages = {103805},
  numpages = {15},
  year = {2018},
  month = {Oct},
  publisher = {American Physical Society},
  doi = {10.1103/PhysRevMaterials.2.103805},
  url = {https://doi.org/10.1103/PhysRevMaterials.2.103805}
}

@Article{Wang2021,
author={Wang, Zifeng
and Ye, Shizhuo
and Wang, Hao
and He, Jin
and Huang, Qijun
and Chang, Sheng},
title={Machine learning method for tight-binding Hamiltonian parameterization from ab-initio band structure},
journal={npj Computational Materials},
year={2021},
month={Jan},
day={25},
volume={7},
number={1},
pages={11},
doi={10.1038/s41524-020-00490-5},
url={https://doi.org/10.1038/s41524-020-00490-5}
}

@article{Seifert1998,
  title = {Self-consistent-charge density-functional tight-binding method for simulations of complex materials properties},
  author = {Elstner, M. and Porezag, D. and Jungnickel, G. and Elsner, J. and Haugk, M. and Frauenheim, Th. and Suhai, S. and Seifert, G.},
  journal = {Phys. Rev. B},
  volume = {58},
  issue = {11},
  pages = {7260--7268},
  numpages = {0},
  year = {1998},
  month = {Sep},
  publisher = {American Physical Society},
  doi = {10.1103/PhysRevB.58.7260},
  url = {https://doi.org/10.1103/PhysRevB.58.7260}
}

@article{Frauenheim2020,
    author = {Hourahine, B. and Aradi, B. and Blum, V. and Bonafé, F. and Buccheri, A. and Camacho, C. and Cevallos, C. and Deshaye, M. Y. and Dumitrică, T. and Dominguez, A. and Ehlert, S. and Elstner, M. and van der Heide, T. and Hermann, J. and Irle, S. and Kranz, J. J. and Köhler, C. and Kowalczyk, T. and Kubař, T. and Lee, I. S. and Lutsker, V. and Maurer, R. J. and Min, S. K. and Mitchell, I. and Negre, C. and Niehaus, T. A. and Niklasson, A. M. N. and Page, A. J. and Pecchia, A. and Penazzi, G. and Persson, M. P. and Řezáč, J. and Sánchez, C. G. and Sternberg, M. and Stöhr, M. and Stuckenberg, F. and Tkatchenko, A. and Yu, V. W.-z. and Frauenheim, T.},
    title = {DFTB+, a software package for efficient approximate density functional theory based atomistic simulations},
    journal = {The Journal of Chemical Physics},
    volume = {152},
    number = {12},
    pages = {124101},
    year = {2020},
    month = {03},
    doi = {10.1063/1.5143190},
    url = {https://doi.org/10.1063/1.5143190},
}

@article{cg,
author = {Daniel, James W.},
title = {Convergence of the conjugate gradient method with computationally convenient modifications},
year = {1967},
issue_date = {July      1967},
publisher = {Springer-Verlag},
address = {Berlin, Heidelberg},
volume = {10},
number = {2},
issn = {0029-599X},
url = {https://doi.org/10.1007/BF02174144},
doi = {10.1007/BF02174144},
journal = {Numer. Math.},
month = jul,
pages = {125–131},
numpages = {7}
}

@article{QE-2017,
  author={P Giannozzi and O Andreussi and T Brumme and O Bunau and M Buongiorno Nardelli
  and M Calandra and R Car and C Cavazzoni and D Ceresoli and M Cococcioni and N Colonna
  and I Carnimeo and A Dal Corso and S de Gironcoli and P Delugas and R A DiStasio Jr and A Ferretti
  and A Floris and G Fratesi and G Fugallo and R Gebauer and U Gerstmann and F Giustino and T Gorni
  and J Jia and M Kawamura and H-Y Ko and A Kokalj and E Küçükbenli and M Lazzeri and M Marsili
  and N Marzari and F Mauri and N L Nguyen and H-V Nguyen and A Otero-de-la-Roza and L Paulatto
  and S Poncé and D Rocca and R Sabatini and B Santra and M Schlipf and A P Seitsonen
  and A Smogunov and I Timrov and T Thonhauser and P Umari and N Vast and X Wu and S Baroni},
  title={Advanced capabilities for materials modelling with QUANTUM ESPRESSO},
  journal={Journal of Physics: Condensed Matter},
  volume={29},
  number={46},
  pages={465901},
  url={https://doi.org/10.1088/1361-648X/aa8f79},
  year={2017},
}

@article{QE-2009,
	Author = {Paolo Giannozzi and Stefano Baroni and Nicola Bonini and Matteo Calandra and Roberto Car
 and Carlo Cavazzoni and Davide Ceresoli and Guido L Chiarotti and Matteo Cococcioni and Ismaila Dabo
 and Andrea {Dal Corso} and Stefano de Gironcoli and Stefano Fabris and Guido Fratesi and Ralph Gebauer
 and Uwe Gerstmann and Christos Gougoussis and Anton Kokalj and Michele Lazzeri and Layla Martin-Samos
 and Nicola Marzari and Francesco Mauri and Riccardo Mazzarello and Stefano Paolini and Alfredo Pasquarello
 and Lorenzo Paulatto and Carlo Sbraccia and Sandro Scandolo and Gabriele Sclauzero and Ari P Seitsonen
 and Alexander Smogunov and Paolo Umari and Renata M Wentzcovitch},
	Journal = {Journal of Physics: Condensed Matter},
	Number = {39},
	Pages = {395502},
	Title = {QUANTUM ESPRESSO: a modular and open-source software project for quantum simulations of materials},
	Url = {https://doi.org/10.1088/0953-8984/21/39/395502},
	Volume = {21},
	Year = {2009}}

@article{doi:10.1063/5.0005082,
 author = {Giannozzi,Paolo  and Baseggio,Oscar  and Bonfà,Pietro  and Brunato,Davide  and Car,Roberto  and Carnimeo,Ivan  
           and Cavazzoni,Carlo  and de Gironcoli,Stefano  and Delugas,Pietro  and Ferrari Ruffino,Fabrizio  and Ferretti,Andrea  
           and Marzari,Nicola  and Timrov,Iurii  and Urru,Andrea  and Baroni,Stefano },
 title = {Quantum ESPRESSO toward the exascale},
 journal = {The Journal of Chemical Physics},
 volume = {152},
 number = {15},
 pages = {154105},
 year = {2020},
 doi = {10.1063/5.0005082},
 URL = {https://doi.org/10.1063/5.0005082},
}

@article{pslibrary,
title = {Pseudopotentials periodic table: From {H} to {Pu}},
journal = {Computational Materials Science},
volume = {95},
pages = {337-350},
year = {2014},
doi = {https://doi.org/10.1016/j.commatsci.2014.07.043},
url = {https://doi.org/10.1016/j.commatsci.2014.07.043},
author = {Andrea {Dal Corso}}
}

@article{boltztrap2,
   title={BoltzTraP2, a program for interpolating band structures and calculating semi-classical transport coefficients},
   volume={231},
   url={http://dx.doi.org/10.1016/j.cpc.2018.05.010},
   DOI={10.1016/j.cpc.2018.05.010},
   journal={Computer Physics Communications},
   publisher={Elsevier BV},
   author={Madsen, Georg K.H. and Carrete, Jesús and Verstraete, Matthieu J.},
   year={2018},
   month=oct, pages={140–145}
}

@article{Pickett1988SmoothFI,
  title = {Smooth Fourier interpolation of periodic functions},
  author = {Pickett, Warren E. and Krakauer, Henry and Allen, Philip B.},
  journal = {Phys. Rev. B},
  volume = {38},
  issue = {4},
  pages = {2721--2726},
  numpages = {0},
  year = {1988},
  month = {Aug},
  publisher = {American Physical Society},
  doi = {10.1103/PhysRevB.38.2721},
  url = {https://doi.org/10.1103/PhysRevB.38.2721}
}

@article{symwannier90,
  title = {Symmetry-adapted Wannier functions in the maximal localization procedure},
  author = {Sakuma, R.},
  journal = {Phys. Rev. B},
  volume = {87},
  issue = {23},
  pages = {235109},
  numpages = {8},
  year = {2013},
  month = {Jun},
  publisher = {American Physical Society},
  doi = {10.1103/PhysRevB.87.235109},
  url = {https://doi.org/10.1103/PhysRevB.87.235109}
}

@incollection{epubdata,
  title={Dataset for "Optimization of ab-initio based tight-binding models"},
  year={2025},
  author={H. Dick},
  doi={10.4119/unibi/3006317},
  url={https://doi.org/10.4119/unibi/3006317}
}

\end{document}